\newcommand{\df}{{\mbox{\rm d}}}
\documentstyle[eqsecnum,aps,epsf]{revtex}

\draft
\preprint{\ }

\begin{document}
\title{Black Holes in Non-flat Backgrounds:\\
the Schwarzschild Black Hole in the Einstein Universe}
\author{K. Rajesh Nayak$^*$,
 M.A.H. MacCallum$^\dagger$ 
  and
 C.V. Vishveshwara$^{ * \S }$ \\ \ \\
$^*$Indian Institute of Astrophysics, \\
 Bangalore 560 034, India \\ 
Email: nayak@iiap.ernet.in\\
 vishu@iiap.ernet.in\\
\ \\
$^{\dagger}$School of Mathematical Sciences,\\
Queen Mary and Westfield College, \\  London, UK\\
Email: M.A.H.MacCallum@qmw.ac.uk\\
\ \\
$^{\S}$Bangalore Association for Science Education (BASE), \\
Bangalore 560 001, India.
}
\date{\today}
\maketitle
\begin{abstract}
As an example of a black hole in a non-flat background a composite
static spacetime is constructed. It comprises a vacuum Schwarzschild
spacetime for the interior of the black hole across whose horizon it is 
matched on to the spacetime of Vaidya 
representing a black hole in the background of the Einstein universe.
The scale length of the
exterior sets a maximum to the black hole mass. To obtain a
non-singular exterior, the Vaidya metric is matched to an Einstein
universe. The behaviour of scalar waves is
studied in this composite model.                                     
\end{abstract}
\pacs{ \ }
\section{Introduction}
        For more than three decades now, black holes have been
investigated in great depth and detail.  However, almost all
these studies have focused on isolated black holes possessing
two basic properties, namely time-independence characterized by
the existence of a timelike Killing vector field, and asymptotic
flatness.  On the other hand, one cannot rule out the important
and, perhaps, realistic situation in which the black hole is
associated with a non-flat background.  This would be the case
if one takes into account the fact that the black hole may
actually  be embedded in the cosmological spacetime or surrounded
by  local mass distributions.  In such situations one or both
of the two basic properties may have to be given up.  
If so, the properties of isolated black holes 
may be modified, completely changed or 
 retained
unaltered.  Black holes in non-flat backgrounds form, therefore,
an important topic.  Very little has been done in this direction.
Some of the issues involved here have been outlined in a recent
article by Vishveshwara \cite{CVV1}.  As has been mentioned in that
article, there may be fundamental questions of concepts and
definitions involved here.  Nevertheless, considerable insight
may be gained by studying specific examples even if they are
not entirely realistic.  In this regard the family of spacetimes
derived by Vaidya \cite{PCV}, which is a special case of Whittaker's
solutions \cite{WITT}, representing in a way black holes in
cosmological backgrounds have been found to be helpful.  Nayak
and Vishveshwara \cite{RNCV} have studied these spacetimes concentrating
on the geometry of the Kerr black hole in the background of the Einstein
universe, which dispenses with asymptotic flatness while preserving
time-symmetry.  In the present paper, we specialize to the simpler
case of the Schwarzschild black hole in the background of the
Einstein universe, which we may call the Vaidya-Einstein-Schwarzschild
(VES) spacetime.  This allows us to study this spacetime in
considerable detail as well as investigate a typical physical
phenomenon, namely the behaviour  of scalar waves
in this spacetime as the  background.

The rest of this paper is organized as follows.  In section II, we
consider the line element of the VES spacetime and the energy-momentum
tensor.  In section III, we match the metric of the VES spacetime
to the Schwarzschild vacuum metric across the black hole surface.
Similarly we match the VES spacetime to the Einstein universe at
large distances.  In section IV, we investigate the behaviour of
scalar waves propagating in this spacetime.  Section V comprises
the concluding remarks. 

\section{The Line Element and the Energy-Momentum Tensor}

        As mentioned earlier, an account of Vaidya's black hole
spacetimes in cosmological spacetimes may be found in references
\cite{PCV} and \cite{RNCV}. By setting the angular momentum to zero in
the Kerr metric we obtain the line element of the Schwarzschild
spacetime in the background of the Einstein universe :
\begin{eqnarray}
\label{VES}
ds^2 &  = & \left( 1 - \frac { 2 m}{ R \tan ( \frac{r}{R})}
\right) dt^2 -
\left( 1 - \frac { 2 m}{ R \tan ( \frac{r}{R})}\right)^{-1}
dr^2   - R^2 \sin^2(\frac{r}{R})
\  \left[ d\theta^2 + \sin^2 \theta
d\phi^2 \right]
\end{eqnarray}
where $m $ is the mass and the coordinates range from 
$ 0 \leq r/R \leq \pi $ , $ 0 \leq \theta \leq \pi  $ and $ 0 \leq 
\phi \leq 2 \pi$.
In the limits $m = 0$  and $R = \infty$, we recover respectively
the Einstein universe and the Schwarzschild spacetime.  The parameter
$R$  is a measure of the cosmological influence on the spacetime.
As the spacetime is static, the black hole is identified as the
surface on which the time-like Killing vector becomes null, {\it i.e.},
$ g_{00} = 0 $  above, which is the static limit and the Killing event
horizon.  The black hole is therefore given by
\begin{equation}
 2 m = R \tan ( \frac{r}{R}). 
\end{equation}
We shall now work out the energy-momentum tensor for this metric.
The components of the Einstein tensor are given by
\begin{equation}
G^1_1 \ = \ G^2_2 \ = \  G^3_3 \ = \ \frac{1}{3} G^0_0 \ = \
\frac{1}{R^2} ( 1 -  \frac{2m}{R \tan(r/R)}).
\end{equation}

        The Einstein field equations, including the cosmological
term $\Lambda$ for generality, although it could equally well be
considered to be included in $\rho$ and $p$ below, are given by
\begin{equation}
R_{ab} - \frac{1}{2}  g_{ab} R \ = \ \kappa T_{ab} + \Lambda g_{ab},
\end{equation}
where $\kappa = \frac{ 8 \pi G}{ c^2} $  and the Latin indices $a,b$
range from $0$ to $3 $ (Greek indices $\mu, \nu = 1-3$).  

The energy-momentum tensor
is taken to be that of a perfect fluid,
\begin{equation}
T_{ab} \ = \  ( \rho + p ) u_a u_b - p g_{ab} ,
\end{equation}
$u^a$  being the static four velocity
\begin{equation}
u^a \ = \  \frac{1}{\sqrt{ g_{00}}} \delta^a_0.
\end{equation}
Then density $\rho$ and pressure $p$  are given by
\begin{eqnarray}
\label{rho}
\rho & = & \frac{ 3 }{ \kappa R^2 } \left(  1 - \frac{ 2 m }
{R \tan(r/R)} \right) - \Lambda/\kappa,  \\
\label{p}
p & = &  \frac{ -1 }{ \kappa R^2 } \left(  1 - \frac{ 2 m }
{R \tan(r/R)} \right) + \Lambda/\kappa .
\end{eqnarray}
The behaviour of $\rho$ and $p$ can be easily ascertained from
the above equations.\\
$\Lambda  > 0$: \\
We find that  $\rho, p \leq 0 $  in some region outside the black hole
violating the weak energy condition.\\
$\Lambda \leq  0$: \\
In this case, $\rho > 0 $ but $p < 0 $ everywhere outside 
the black hole and tends to zero on it.
 However, $\rho + p \geq 0 $ thereby satisfying the 
weak energy condition.  

For convenience, we take $ \Lambda  = 0. $ 
Then
\begin{equation}
\rho + 3 p \ = \ 0. 
\end{equation}
This suggests that the spacetime is a  special case of the solutions obeying
the condition $\rho + 3 p \ = constant$ discussed by Whittaker
\cite{WITT}, and it is easy to check that this is so (it is the case
$B=G=\lambda=0$, $c=-2m$, $\alpha = 1/R$, with the time coordinate scaled so
that $n=1$, in Whittaker's notation).

Thus the behaviour of the energy-momentum tensor is reasonable, since
in the Einstein universe itself we have $p<0$, while $\rho$  and $p$
satisfy the weak energy condition.
\section{Matching to the Schwarzschild vacuum and the Einstein Universe}
        In this section we shall match the VES spacetime to the
Schwarzschild vacuum spacetime on one side and to the Einstein
universe on the other.  The possibility of matching to the
Schwarzschild vacuum at the black hole surface, without a surface
layer or shell, is strongly indicated by the fact that the Einstein
tensor of the VES spacetime goes to zero on the surface.  We shall
show that this is indeed possible.  In order to do this, we will first
write the line element in Kruskal-like coordinates, among which we will
find admissible coordinates in which the matching can be carried out,
so that the requirements \cite{MARS} become simply the continuity of the metric
and its first derivative.

        The Kruskal form of the VES line element is arrived at by
the following transformations. 
\begin{eqnarray}
r^* & = &  \frac{R^2}{ 4 m^2 + R^2}\left( r + 2 m \ln \left[ - 2 m
    \cos(\frac{r}{R} ) + R \sin(\frac{r}{R} ) \right]\right), \\ 
u &=& t - r^*, \quad  v = t + r^*, \\
\widehat{U} & = &  -\exp\left[ -\frac{u}{4m} \frac{ 4 m^2 + R^2}{R^2} \right], \\
\widehat{V} & = & \exp\left[ \frac{v}{4m} \frac{ 4 m^2 + R^2}{R^2}
    \right].
\end{eqnarray}
Then we obtain 
\begin{equation}
\df s^2 = (\frac{4mR^2}{4m^2+R^2})^2 \frac{1}{R\sin (r/R)} e^{-r/2m}
\df \widehat{U} \,\df \widehat{V} - (R\sin (r/R))^2 \df \Omega^2. \label{KURSKL}
\end{equation}
The Kruskal line element for the Schwarzschild vacuum spacetime,
\begin{equation}
\df s^2 = 16m_s^2 \frac{1}{r_s} e^{-r_s/2m_s}
\df \widehat{U} \,\df \widehat{V} - r_s^2 \df \Omega^2,
\end{equation}
may be recovered from equation (\ref{KURSKL}) by the limit $ R =
\infty$. As usual the Kruskal coordinates for the Schwarzschild space
cover the whole maximally extended spacetime and not only the region
where the coordinates $t,\,r$ are valid.
Now we proceed to carry out the matching at the horizons. 

The horizon of the VES metric is at $r=r_0$ where $2m=R\tan
(r_0/R)$. To match to Schwarzschild at the horizon the angular variables part
requires  $2m_s=R\sin(r_0/R)$. Let us use $r^\prime =R\sin (r/R)$ as
the radial variable in the VES region. We
can rescale both the $\widehat{U}$ and $\widehat{V}$ of each of the metrics by constant
factors $4m_s/\sqrt{e}$ and $4mR^2e^{-r_0/4m}/(4m^2+R^2)$ respectively,
giving new coordinates U, V,  to reduce the metrics to the forms,
\begin{equation}
\df s^2 = \frac{1}{r^\prime} e^{(r_0-r)/2m}
\df U \,\df V - (r^\prime)^2 \df \Omega^2,
\end{equation}
\begin{equation}
\df s^2 = \frac{1}{r_s} e^{(2m_s-r_s)/2m_s}
\df U \,\df V - r_s^2 \df \Omega^2.
\end{equation}
Then we see the metric is continuous if we identify $r_s$ and
$r^\prime$ at the future horizon
$U=0,~r'=r_s=2m_s=R\sin(r_0/R),~r=r_0$. To
deal with derivatives, start with
\begin{equation}
UV = (\frac{4mR^2}{4m^2+R^2})^2 e^{-r_0/2m}
 \exp(2{\frac{4m^2+R^2}{4mR^2}}r^\ast) 
\end{equation}
on the VES side,  so that there
\begin{equation}
V = 2(\frac{4mR^2}{4m^2+R^2}) e^{-r_0/2m}
\exp({2\frac{4m^2+R^2}{4mR^2}}r^\ast) \frac{\df r^\ast}{\df U}
\end{equation}
and  therefore
\begin{eqnarray*}
\frac{\df r^\prime}{\df U} &=&
\frac{\df r^\prime}{\df r}\quad\quad\quad \frac{\df r}{\df r^\ast}
\quad\quad\quad\quad\quad \frac{\df r^\ast}{\df U}\\
&=&
\cos(r/R)~ (1-\frac{2m}{R\tan(r/R)})~ Ve^{r_0/2m}\frac{4m^2+R^2}{8mR^2}
\exp(-2{\frac{4m^2+R^2}{4mR^2}}r^\ast).
\end{eqnarray*}
As $r \rightarrow r_0$ the product
$(1-\frac{2m}{R\tan(r/R)})\exp(-2{\frac{4m^2+R^2}{4mR^2}}r^\ast)$ approaches 
$e^{-r_0/2m}/2m_s$ and we get
\begin{eqnarray*}
\frac{\df r^\prime}{\df U} &=&\cos(r_0/R) V \frac{4m^2+R^2}{16mm_sR^2}\\
&=& V\frac{R}{\sqrt{4m^2+R^2}}\frac{4m^2+R^2}{16mm_sR^2}\\
&=& V\frac{\sqrt{4m^2+R^2}}{2mR}\frac{1}{8m_s} = \frac{V}{16m_s^2}
\end{eqnarray*}
which is obviously the same as for $\displaystyle{\frac{\df r_s}{\df U}}$ in the
Schwarzschild metric. Now the derivatives of the metric coefficients
will match if
\begin{equation}
\frac{1}{2m}\frac{\df r}{\df r^\prime}=\frac{1}{2m_s}=\frac{1}{R\sin(r_0/R)}
\end{equation}
at the horizon, but $\displaystyle{\frac{\df r}{\df r^\prime}=1/\cos(r/R)}$ 
and consequently
\begin{equation}
\frac{1}{2m}\frac{\df r}{\df
r^\prime}=\frac{1}{2m\cos(r_0/R)}=\frac{1}{R\sin(r_0/R)}
\end{equation}
because at the horizon $2m=R\tan(r_0/R)$.  This completes the matching at the 
future horizon. Clearly a similar matching with the roles of $U$ and
$V$ reversed applies at the past horizon in a Kruskal picture.

        We may note that matching the metric component $g_{33}$ yields
the relation between the Schwarzschild vacuum mass $m_s$ and the VES
mass $m$,
\begin{equation}
m = m_s \left[  1 - \left( \frac{ 2 m_s }{R} \right) ^2 \right]^{-\frac{1}{2}}.
\end{equation}
 This clearly exhibits the influence of the cosmological matter distribution
on the bare black hole mass.  Figure 1 shows plots of $m$ as a function
of $m_s$ for different values of $R$. We note that $2m_s \leq R$, so
the length scale in the exterior puts a bound on the black hole mass,
in a way which may be analogous with the bound found in \cite{NAK} for
the mass of a black hole in an Einstein space.

\begin{figure}[h]
\centerline
{\epsfbox{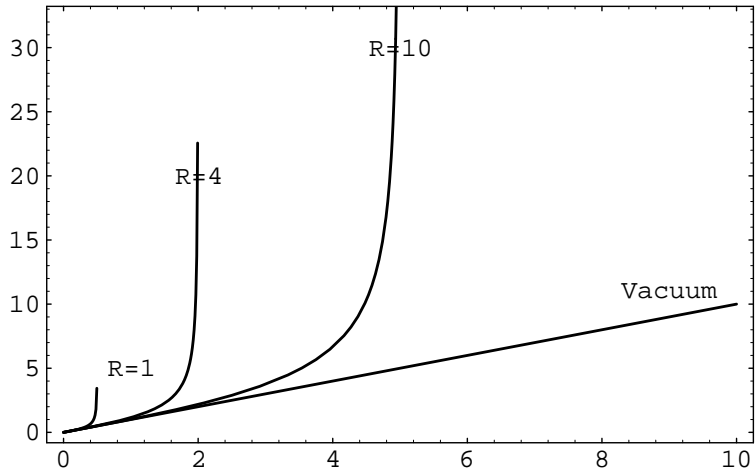}}
\caption{ Plot of $m$  as a function of $m_s$ for different values of
$R$
 } 
\end{figure}

It is also worth emphasizing that a consequence of this matching is
that all the horizon properties, such as the specific gravity, are
necessarily the same as those of the usual Schwarzschild black
hole. Whether this is reassuring or disappointing is a matter of
opinion. It does not imply, however,  that properties which depend on the
behaviour in the exterior region, such as the behaviour of waves,
will be the same.

To investigate such behaviours we need a well-behaved non-vacuum
exterior. Unfortunately, the formulae (\ref{rho}) and (\ref{p}) show
that the energy density and pressure of the VES spacetime blow up as
$r/R \rightarrow \pi$
and this is in fact a naked singularity. To remove it we try to match
to the Einstein universe (which, remember, is a limit of
(\ref{VES})). It is easy to see that the best hope of doing so without
a surface layer is at $r/R= \pi/2$, where we could match both the
angular part of the metric and its derivative. In fact, at this radius
%
the VES line element reduces to 
\begin{equation}
\df s^2 \ = \  \df t^2 \ - \ \df r^2 \ - \  R^2 \sin^2\left( \frac{r}{R} 
\right) \df \Omega^2
\end{equation} 
which is the line element of the Einstein universe.  The metric
components of the two spacetimes automatically match, without any
change of coordinates, and the first
derivative of the angular parts on both sides vanishes.  But the
first derivative of the $tt$-parts is discontinuous thereby giving
rise to a surface
distribution of matter.  The components of the corresponding energy-momentum
tensor may be computed following Mars and Senovilla \cite{MARS}.  We find that
this leads to a trace free tensor.

More specifically, the jump in the fundamental form of the
$r=constant$ surfaces is
\begin{equation}
\label{Kjump}
[K_{tt}] = -m/R^2
\end{equation}
and the non-zero  components of the $\delta$-function parts of the
curvature and Ricci tensor are given by
\begin{equation}
\label{Rjump}
Q^r{}_{ttr} = -m/R^2, ~~R_{tt} = R_{rr}= m/R^2.
\end{equation}
Such a layer might be interpreted as a domain wall.

        We  now have a composite model consisting of a vacuum Schwarzschild
black hole matched onto the VES spacetime which is itself matched to the
Einstein universe.

\section{Scalar Waves}

      In the last section we constructed a model for a black hole
in a non-flat background. The interior of the black hole consists 
of the Schwarzschild vacuum.  The exterior is the VES spacetime matched
on to the Einstein universe.  One can explore black hole physics in
the exterior and compare it with the effects one encounters in the
case of the usual isolated black holes.  As an example of such possible
studies, we shall consider some properties of scalar waves propagating in 
this spacetime. Other phenomena occurring in this spacetime, such
as the classical tests of general relativity  and the  geodesics,
have been investigated by Ramachandra and
Vishveshwara \cite{RAMCV}.

      Because of the time and spherical symmetries of the spacetime, the
scalar wave function may be decomposed as
\begin{equation}
\psi \ = \ e^{i \omega t}{\cal R}(r) Y_l^{ \ m} ( \theta , \phi).
\end{equation}

      The limits of the radial coordinate are given by $ R \tan(r/R) =2m $
 to $(r/R) = \pi$ with
the VES spacetime extending from $  R \tan(r/R) =2m $  to 
$ r/R = \pi/2 $  and the Einstein
universe from $r/R = \pi/2$ to $\pi$.  We set the radial function
\begin{equation}
{\cal R} (r) \ = \  \frac{u(r)}{ R \sin(r/R)}
\end{equation}
and define
\begin{equation}
\df r^* \ = \ \frac{dr}{ 1 - \frac{ 2m }{R \tan(r/R)}}.
\end{equation}
Then we obtain the Schr\"{o}dinger equation governing the radial function 
\begin{equation}
\frac{ d^2 u}{d {r^\ast}^2} + \left[ \omega^2 - V(r)
\right] u  \ = \ 0.
\end{equation}
The effective potential that controls the propagation of the scalar
waves is given by
\begin{eqnarray}
V(r) \ = \  \left( 1 - \frac{ 2 m } { R \tan( \frac{r}{R})}
\right) \left[ \frac{ l(l+1) } { R^2 \sin^2 ( \frac{r}{R}) }
+ \frac{ 2 m }
 { R^3 \sin^2 ( \frac{r}{R})\tan( \frac{r}{R}) } \right.
\nonumber \\
\left.
- \frac{1}{R^2} \left( 1 - \frac{ 2 m }{ R \tan
( \frac{r}{R} ) }\right) \right].
\end{eqnarray}

      We shall now discuss a few aspects of the behaviour of the
scalar waves as reflected by the nature of the effective potential.

\begin{figure}[h]
\centerline
{\epsfbox{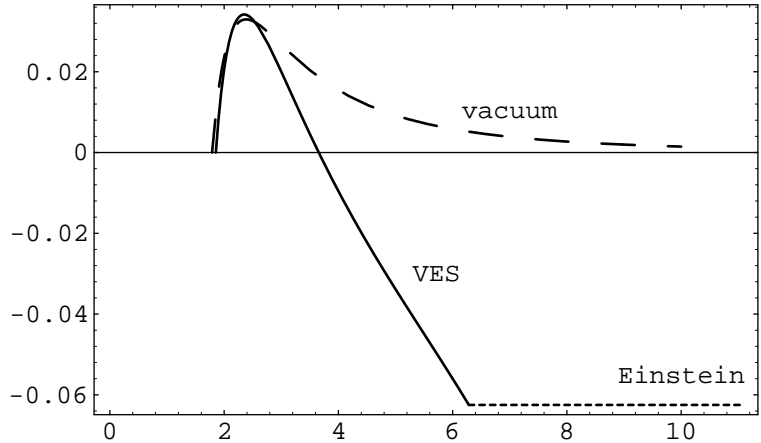}}
\caption{ Plot of effective potential $V(r)$ of VES spacetime 
for $R/m=4$ and $l=0$.
For comparison the effective potentials of the vacuum
Schwarzschild spacetime (dashed line) and the Einstein
universe (dotted line) are also shown. 
 } 
\end{figure}
      We have drawn $V(r)$ in fig.2 for $l=0$. The figure shows
the corresponding effective potential for the vacuum Schwarzschild
exterior also which can be obtained by setting $R=\infty$. Both curves
start from zero at the black hole and go through a maximum. Thus both
potentials possess potential barriers. As in the case of the Schwarzschild
vacuum, now too waves can be reflected at the barrier while the
transmitted
part is absorbed by the black hole. On the other hand, whereas the
vacuum potential goes asymptotically to zero, in the present case the
potential becomes negative at $ r/R = \pi/2$  and continues as a
constant, {\it i.e} $-\frac{1}{R^2}$, in the
Einstein universe up to $r/R=\pi$. The fact that the effective potential is
negative as above raises the possibility of $\omega^2$ being negative as
well. This would be equivalent to $\omega$ being imaginary thereby giving
rise to exponential growth in time of the scalar wave function. This
would mean instability of the  model spacetime against scalar perturbations.
However, one can see that negative values of $\omega^2$ are ruled out by
the boundary condition at $r/R= \pi$. In the Einstein universe sector the
Schr\"{o}dinger equation reduces to
\begin{equation}
\frac{ \df^2 u}{ \df  r^{*2}} + \left( \omega^2 + \frac{1}{R^2}
\right) u \ = \ 0.
\end{equation}
We note that in the Einstein universe, we have $r^*=r$.  Furthermore,
since $ {\cal R} = \frac{u(r)}{ R \sin(r/R)} $, the function $u \sim
\sin\left[\left(\omega^2 + \frac{1}{R^2} \right)^{\frac12} r \right] $
has to go to zero faster than $ \sin(r/R)$ at $r/R=\pi$. This boundary
condition requires that $|(R^2\omega^2+1)^{\frac12}|$ be an integer
greater than 1, and thence that $\omega^2$ be positive. Therefore the
spacetime is stable against scalar perturbations.

This is true in the case of vacuum Schwarzschild spacetime as well
as the Einstein universe.
However, the stability against gravitational perturbations is a
different matter altogether. Whereas the Schwarzschild  vacuum
exterior is stable, the Einstein universe is not \cite{EDIG}.
Whether the combination of the two spacetimes is stable, unstable 
or conditionally stable is an intriguing open question.

\begin{figure}[h]
\centerline
{\epsfbox{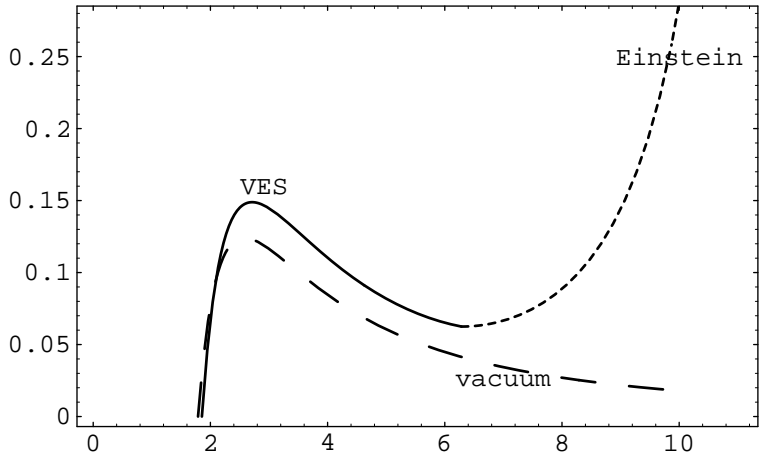}}
\caption{ Plot of effective potential $V(r)$ for $R/m=4$ and $l=1$.
 } 
\end{figure}
For $ l >0 $, the equivalent potential has the additional term 
$ \frac{ l (l+1)}{ R^2 \sin^2 (r/R)}$.

We sketch $V(r)$ for $ l=1$ in figure 3. Once again the potential
goes to zero at the black hole and possesses a barrier region. The
additional term goes to infinity at $(r/R) = \pi $ thereby behaving like a 
centrifugal barrier commonly encountered in the scattering phenomenon.
The radial function is exponential wherever the value of $V(r)$ is
greater than $ \omega^2$ and is a running wave when $\omega^2 > V(r)$. Details of such
solutions can be studied easily.
\section{Concluding Remarks}                
      The motivation for the present work stems from the need for 
detailed study of black holes in non-flat backgrounds in comparison and
contrast to isolated black holes. A comprehensive investigation of
this problem would be a formidable task indeed. We have confined ourselves
in this paper to  a specific example that relaxes the condition of 
asymptotic flatness while preserving time-symmetry. The starting point
here is the static black hole in the Einstein universe which belongs
to the family of solutions presented by Vaidya. In this spacetime
the black hole is well defined as the Killing horizon. However, the
nature of the interior of the black hole is not entirely clear.
Furthermore, it is not obvious {\it a priori} whether the exterior
can be matched smoothly to the Schwarzschild vacuum across the black hole
surface. We have shown that this is possible by carrying out this matching
using Kruskal coordinates in the two regions. Similarly we have 
matched the spacetime to the Einstein universe at the other end. This
provides a composite model of a black hole in a non-flat background.

      In the spacetime considered above, different phenomena may be 
studied and compared to their counterparts in the gravitational field
of an isolated Schwarzschild black hole.
      As an example, we have briefly discussed the behaviour of scalar
waves. The spacetime being considered proves to be stable against 
scalar perturbations as is the Schwarzschild vacuum exterior. This is
true of the Einstein universe as well. However, whereas the Schwarzschild
spacetime is stable against gravitational perturbations, the Einstein
universe is not.  It would be quite interesting to see whether the
spacetime we have considered, which involves both of the above ones, is
gravitationally stable or not.
Even if the model presented here is unrealistic, it should
provide a testing ground for investigating external influences on
the otherwise  isolated black holes.
\subsection*{Acknowledgements}
One of us (C.V.V) would like to thank Prof. M.A.H. Mac Callum and
his colleagues for hospitality during his visit to Queen Mary and
Westfield College. This visit was made possible by a Visiting
Fellowship grant from the UK Engineering and Physical Sciences Research
Council, grant number GR/L 79724, which partially supported the
research reported in this paper.

\end{document}